\documentclass[10pt,letterpaper,twocolumn]{article} 

\usepackage{ol2}
\usepackage[draft]{hyperref}
\usepackage{amsmath}

\begin{document}

\twocolumn[ 

\title{Low-loss directional cloaks without superluminal velocity or magnetic response}


\author{Yaroslav Urzhumov$^{1*}$ and David R. Smith$^{1,2}$}

\address{
$^1$Department of Electrical and Computer Engineering, Duke University \\ 130 Hudson Hall, Box 90271, Durham, N.C. 27708, USA
\\
$^2$Center for Metamaterials and Integrated Plasmonics, Duke University \\ FCIEMAS 2527, Box 90291, Durham, N.C. 27708, USA
\\
$^*$Corresponding author: yaroslav.urzhumov@duke.edu
}

\begin{abstract}
The possibility of making an optically large (many wavelengths in diameter) object appear invisible has been a subject of many recent studies. Exact invisibility scenarios for large (relative to the wavelength) objects involve (meta)materials with superluminal phase velocity (refractive index less than unity) and/or magnetic response. We introduce a new approximation applicable to certain device geometries in the eikonal limit: piecewise-uniform scaling of the refractive index. This transformation preserves the ray trajectories, but leads to a uniform phase delay.
We show how to take advantage of phase delays 
to achieve a limited (directional and wavelength-dependent) form of invisibility that does not require loss-ridden (meta)materials with superluminal phase velocities.
\end{abstract}


 ] 

The attainability of optical invisibility devices (hereafter, cloaks) has been a subject of active research and some debate in the recent years, immediately following the demonstration of a microwave cloak~\cite{schurig_smith06} based on the transformation optics~\cite{pendry_smith06} (TO) methodology. Recent progress in this area can be attributed to the development of electromagnetic (EM) metamaterials with unusual properties such as strong magnetic response and large, controlled anisotropy across the spectrum, from radio~\cite{urzhumov_padilla_smith_prb12,huang_zhang12} to terahertz~\cite{chen_averitt06} and even optical~\cite{sarychev_shalaev07} frequencies.

While TO is not the only methodology proposed for invisibility~\cite{alu_engheta_prl08,tretyakov_simovski09}, so far it is the only known path to cloaking of very large (relative to free-space wavelength $\lambda_0$) objects.
To date, proposed and demonstrated TO cloaks based on passive, linear, reciprocal media require at least three~\cite{pendry_smith06,cai_milton07,urzhumov_smith_njp10,urzhumov_pendry11,urzhumov_landy_smith_jap12} EM properties that are hard to find in natural substances. One key ingredient is the large and precisely controlled {\it anisotropy} of the refractive index~\cite{pendry_smith06,urzhumov_smith_njp10} (RI). Implementations of TO recipes generally require anisotropic materials, with the exception of conformal maps (CM). The usefulness of CMs for cloaking designs is fundamentally limited by several factors, including the preservation of the conformal modulus of the transformed domain, lack of rotational symmetry~\cite{urzhumov_pendry11,urzhumov_landy_smith_jap12}, and the absence of useful CMs in three dimensions. In 2D, lacking radially symmetric CMs, the CM subset of TO can at best offer cloaking for one propagation direction, as was demonstrated by Leonhardt~\cite{leonhardt06}. We refer to such devices as {\it directional cloaks} here. CM cloaks also suffer from other issues, such as the impossibility to conformally map a finite-area domain onto a smaller domain, which leads to an inevitable impedance mismatch at the finite radius cutoff~\cite{urzhumov_pendry11,urzhumov_landy_smith_jap12}; such issues are easily avoided in non-conformal TO scenarios with anisotropic media.

The second and more fundamental cloaking requirement is {\it superluminal phase velocity}, or RI $n<1$. Generally, the requirement is $n<n_0$, where $n_0$ is RI of immersion medium~\cite{zhang_barbastathis11}; we consider only free space cloaking, $n_0=1$. This requirement persists even in CM-based cloaking scenarios~\cite{leonhardt06,landy_kundtz_smith_prl10,urzhumov_pendry11,urzhumov_landy_smith_jap12}.
To implement $n<1$, at least one of the principal values of the
$\epsilon$ and $\mu$ tensors of a non-bianisotropic medium would need to be less than unity. Transparent media with $n<1$ are necessarily dispersive --- otherwise they could be used to transmit information-carrying signals with superluminal speed, a clear contradiction with special relativistic causality. By virtue of Kramers-Kronig relations, one may deduce that effective media with $\epsilon<1$ (or $\mu<1$) must be lossy because of their dispersion. Photonic band gap media operating beyond the effective medium regime in a high-order Bloch band have been proposed as superluminal velocity substrates for lossless, all-dielectric TO devices~\cite{urzhumov_smith_prl10}; however, they still suffer from the dispersion and impedance mismatch limitations.

The third exotic property needed for TO invisibility designs is {\it magnetic response}. In 3D, a non-unity magnetic permeability ($\mu$) is needed both for impedance matching as well as birefringence suppression in anisotropic media. In 2D, for in-plane wave propagation, one may treat the decoupled TE and TM polarizations separately, requiring the management of only a subset of the constitutive tensor elements. For TM waves, the magnetic permeability $\mu_{zz}$ is a scalar. In the short-wavelength (eikonal) limit, one may further simplify the material properties of a TM-polarization cloak and thereby eliminate magnetic response~\cite{cai_milton07,urzhumov_pendry11}.
Elimination of non-unity $\mu$ from cloaking
is valuable particularly for shorter wavelengths, where traditional magnetic materials have weak response and artificial magnetic metamaterials lead to a strongly dispersive and lossy $\mu$~\cite{sarychev_shalaev07}.
Impedance mismatch can be eliminated or reduced using an immersion medium with a properly chosen RI~\cite{zhang_barbastathis11}; however, this technique is not useful for practical applications requiring cloaking in free space or low-density fluids.


In this Letter, we concentrate on the simplified case of directional cloaks~\cite{leonhardt06}, further restricted to 2D in-plane propagation. Our main goal is to show that superluminal phase velocity can be eliminated using a variant of the eikonal approximation, namely, the scaling of RI. Additionally, we show that the Fabry-Perot resonance (FPR) of the entire structure can be used to avoid the requirement of non-unity $\mu$ from directional TM-wave cloaks.


\begin{figure}[htb]
\centering
\begin{tabular}{cc}
\includegraphics[width=0.45\columnwidth]{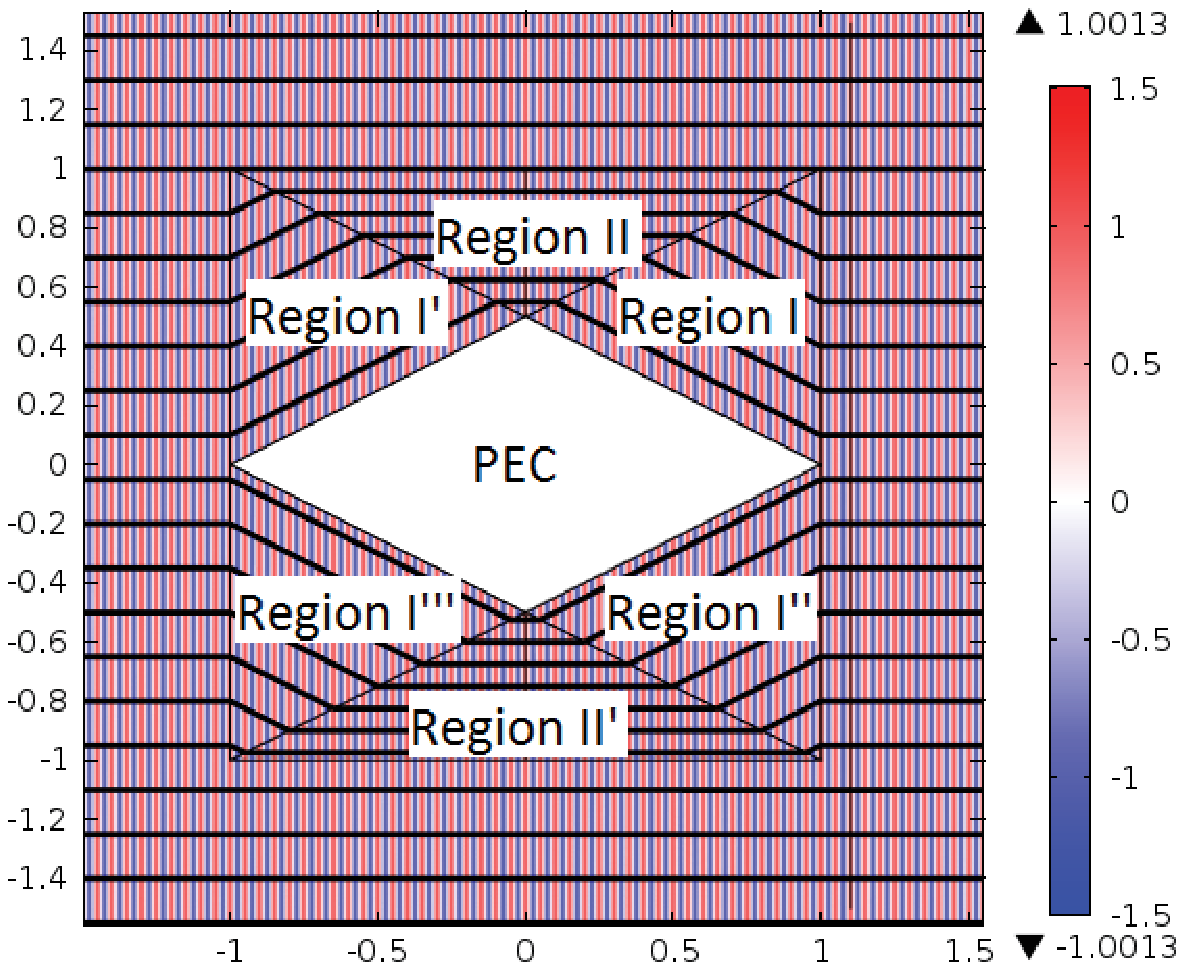}&
\includegraphics[width=0.45\columnwidth]{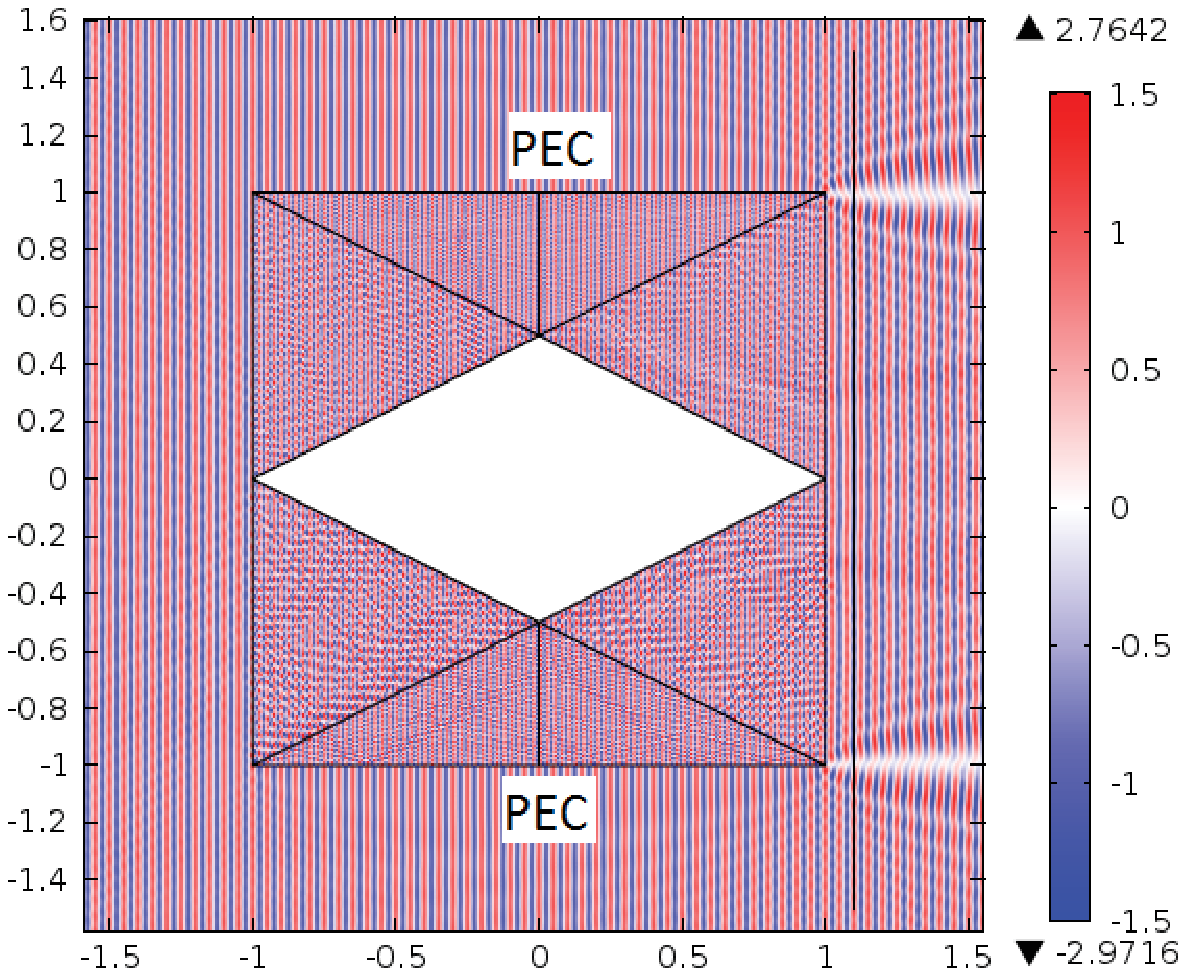}\\
(a)&(b)\\
\includegraphics[width=0.45\columnwidth]{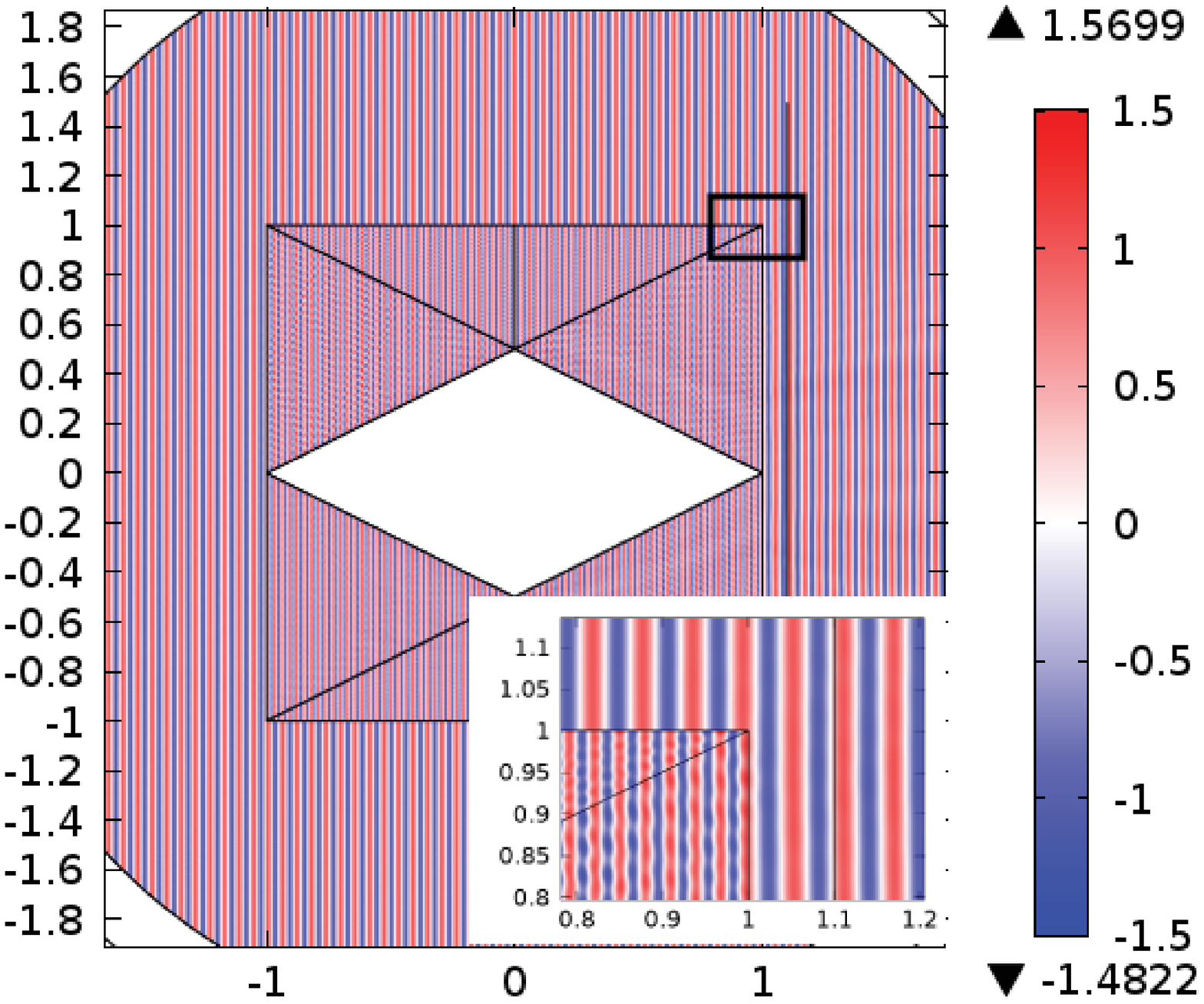}&
\includegraphics[width=0.45\columnwidth]{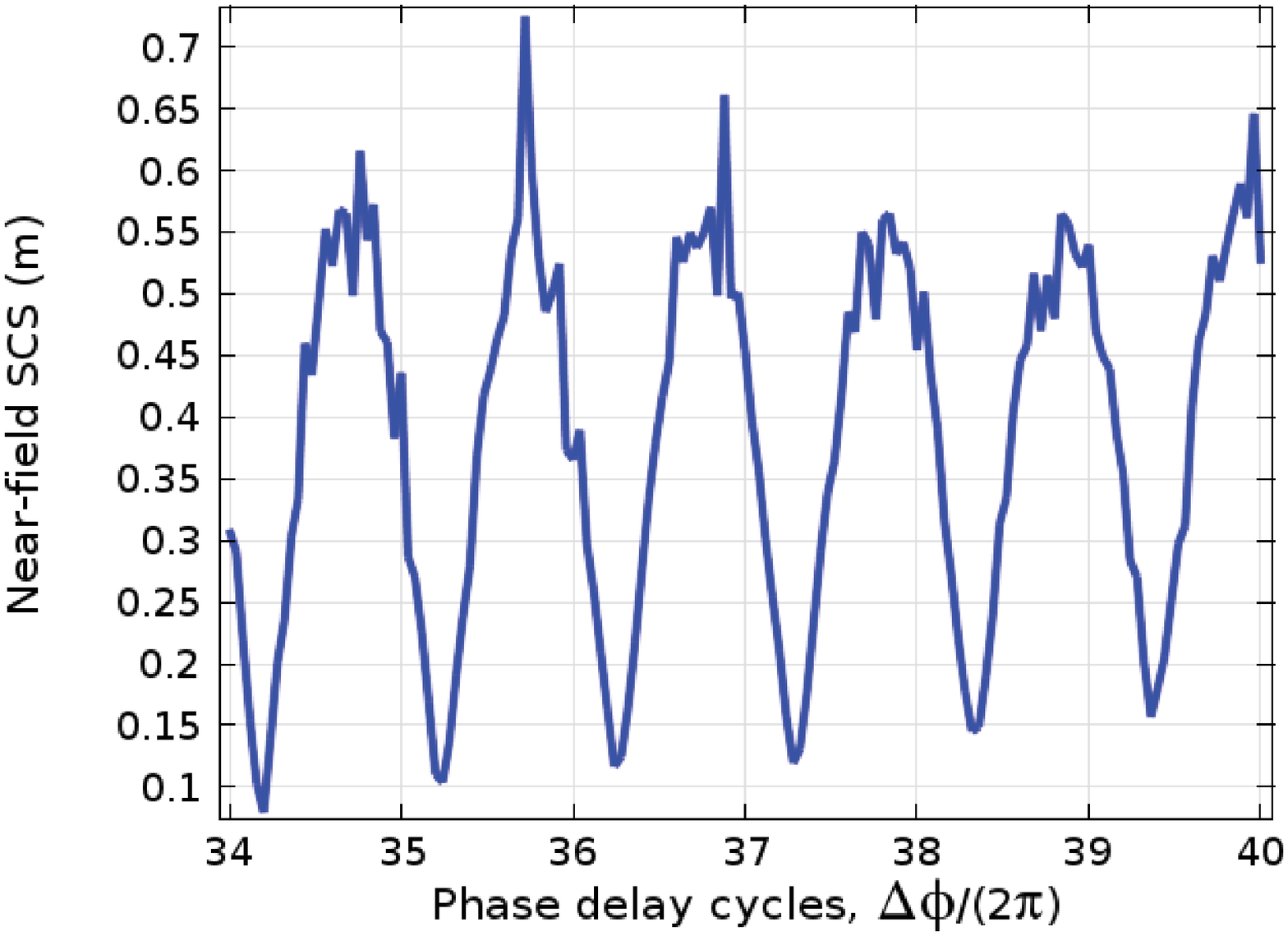}\\
(c)&(d)\\
\end{tabular}
\caption{(color online). The effect of RI scaling in a TM-polarization cloak with $a=1$~m, $b/a=0.5$ and $B/a=1$:
(a) Field plot ($H_z$) at $\lambda_0=(2a)/40$; no RI scaling ($f_i=1$). Black lines are EM flux streamlines, coincident with ray traces in the eikonal limit.
(b) Same as (a) but with RI scaling factor $f_i=2$ (sufficient to make $\epsilon_{1,2}\ge 1$); $\lambda_0=(2a)/40$.
(c) Same as (b) ($f_i=2$), but at $\lambda_0=(2a)/34.2$. 
(d) Measure of visibility, $\sigma_{NF}$, vs frequency in the index-scaled cloak with $f_i=2$; abscissa is
$\Delta\phi/(2\pi)=k_0 (2a)(f_i-1)/(2\pi)$ (number of phase delay cycles per cloak length $2a$).
}
\label{fig:cloak_index_scaled}
\end{figure}

Our cloak geometry is shown in Fig.~\ref{fig:cloak_index_scaled}(a). The choice of geometry is not unique; the only requirement is that all exterior boundaries of the cloak are normal and parallel to the incidence direction.
The 2D Helmholtz equation is modeled using the finite element method with COMSOL RF solver.
The diamond-shaped arbitrary-medium object of length $2a=2$~m and height $2b=1$~m centered at the origin of Cartesian coordinates is coated by a perfectly conducting (PEC) sheet. The PEC-covered diamond is surrounded by a cloak with rectangular exterior boundary (length $2a$ and height $2B=2a$).

The cloak consists of six homogeneous triangular pieces of material, labeled Region I, II, $\rm I'-I'''$ and $\rm II'$ in Fig.~\ref{fig:cloak_index_scaled}(a). Using the standard TO prescription~\cite{pendry_smith06}, the material properties in regions I, II are given by
\begin{eqnarray}
\epsilon^{\rm I}_{xx}=1, \; \epsilon^{\rm I}_{xy}=\epsilon^{\rm I}_{yx}=r, \; \epsilon^{\rm I}_{yy}=1+r^2, \; \mu^{\rm I}_{zz}=1;\nonumber \\
\epsilon^{\rm II}_{xx}=\chi, \; \epsilon^{\rm II}_{xy}=\epsilon^{\rm II}_{yx}=0, \; \epsilon^{\rm II}_{yy}=1/\chi, \; \mu^{\rm II}_{zz}=\chi;
\label{eqn:materials}
\end{eqnarray}
where $\chi=B/(B-b)>1$ and $r=b/a$.
Regions $\rm I'-I'''$ and $\rm II'$ are filled with a mirror-imaged material from the region I and II.
These pieces implement a piecewise-linear, continuous transformation,
\begin{eqnarray}
{\rm I}: x'=x, \; y'=y-r x - b, \nonumber \\
{\rm II}: x'=x, \; y'=\chi (y-b),
\label{eqn:transformations}
\end{eqnarray}
and likewise in regions $\rm I'-I'''$ and $\rm II'$,
the result of which is the appearance of the diamond-shaped PEC object as a flat, zero-thickness, PEC sheet.
In regions $\rm I$ and $\rm I'$, the transformation is area-preserving; thus, $\mu^{\rm I}_{zz}=1$.
The principal values of the dielectric tensors are $\epsilon^{\rm I}_{1,2}=1+\frac{r^2}{2}\pm r\sqrt{1+\frac{r^2}{4}}$; note that $\epsilon^{\rm I}_2<1$. In regions $\rm II$ and $\rm II'$, the transformation increases the area of the cloak inwards, thus allowing the object to shrink. Inevitably, $\mu^{\rm II}_{zz}>1$. In the same domains, the principal values of the dielectric tensors are $\epsilon^{\rm II}_{1,2}=\chi^{\pm 1}$; notably, $\epsilon^{\rm II}_2=1/\chi<1$. Next, we propose a scaling transformation that ensures $\epsilon^{\rm I,II}_{1,2}\ge 1$.


To mitigate the need for RI or dielectric constant of less than unity, we introduce an
approximation that preserves ray trajectories without necessarily maintaining the correct phase.
This control is achieved by a uniform scaling of RI throughout the entire device, while simultaneously preserving the wave impedance. The exterior index remains $n_0=1$ in our procedure, which is different from the immersion approach~\cite{zhang_barbastathis11}.
This procedure creates two issues: (a) RI discontinuity on the surface of the device that causes unwanted reflections, and (b) phase discontinuity between the rays traveling inside and outside the cloak.

To solve the issue (a), we have chosen our device geometry and function such that the RI discontinuity is perpendicular to the wave incidence. At normal incidence, only the wave impedance affects the reflection coefficient, and RI jumps are invisible. This technique, of course, would not be applicable to omnidirectional cloaks, which are beyond the scope of this Letter. For the cloak shown in Fig.~\ref{fig:cloak_index_scaled}(a), the RI needs to be scaled up by at least a factor of $f_i=\chi=2$ to ensure $\epsilon^{\rm I,II}_{1,2}\ge 1$.

Issue (b) causes interference between the rays passing close to the upper and lower boundaries of the cloak (at $y=\pm B$). The interference along those boundaries can be suppressed by placing a pair of geometrically thin PEC sheets at $y=\pm B$ spanning the entire cloak length, from $x=-a$ to $+a$. The result is seen in Fig.~\ref{fig:cloak_index_scaled}(b). From one view angle, these PEC sheets are themselves invisible to TM waves in the eikonal limit, thereby preserving the cloak function.

Even with the PEC shields, there is still significant interference past the cloak (Fig.~\ref{fig:cloak_index_scaled}(c)). This interference is strongly suppressed at discrete frequencies, 
as seen from Fig.~\ref{fig:cloak_index_scaled}(c,d). 
In Fig.~\ref{fig:cloak_index_scaled}(d), we plot a measure of visibility defined as the norm of the magnetic field perturbation integrated over a line in the shadow of the cloak:
$\sigma_{NF}=\int \left|\left|\frac{H_z}{H_{z0}}\right|^2-1\right| ds$,
where $H_{z0}$ is the amplitude of the incident wave. In our calculations, the line of integration is placed at $x=1.1a$, and its length is limited to $3B$. The quantity $\sigma_{NF}$ has the units of length (in 2D), and therefore can be regarded as a near-field scattering width. Evident from Fig.~\ref{fig:cloak_index_scaled}(b,c,d), the minima of $\sigma_{NF}$ are good indicators of scattering suppression. One would expect that these minima should correspond to frequencies at which the phase delay between the inside and outside rays is an integer multiple of $2\pi$; however the presence of diffracted waves causes the minima to be detuned from those positions.


\begin{figure}[htb]
\centering
\begin{tabular}{cc}
\includegraphics[width=0.45\columnwidth]{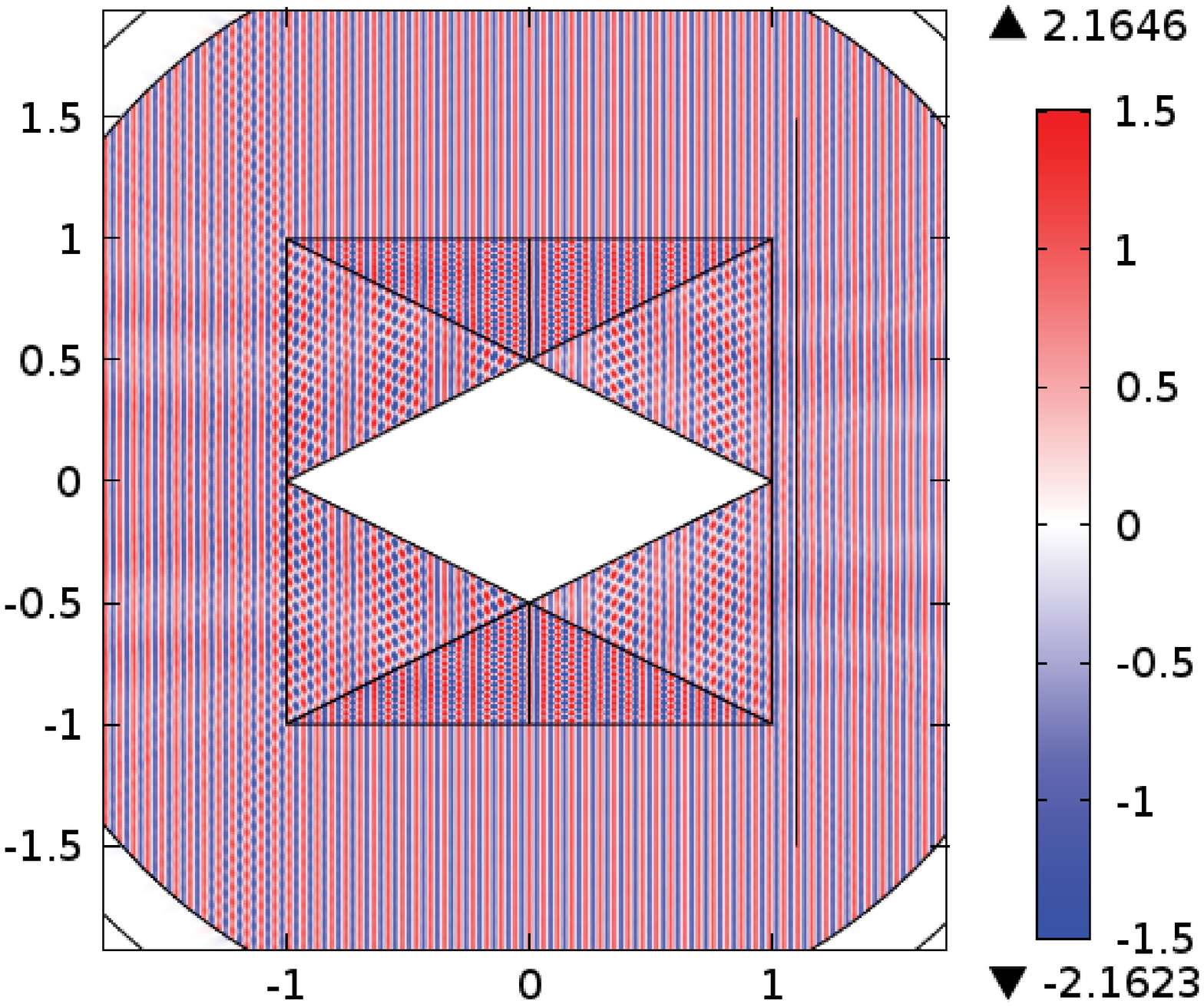}&
\includegraphics[width=0.45\columnwidth]{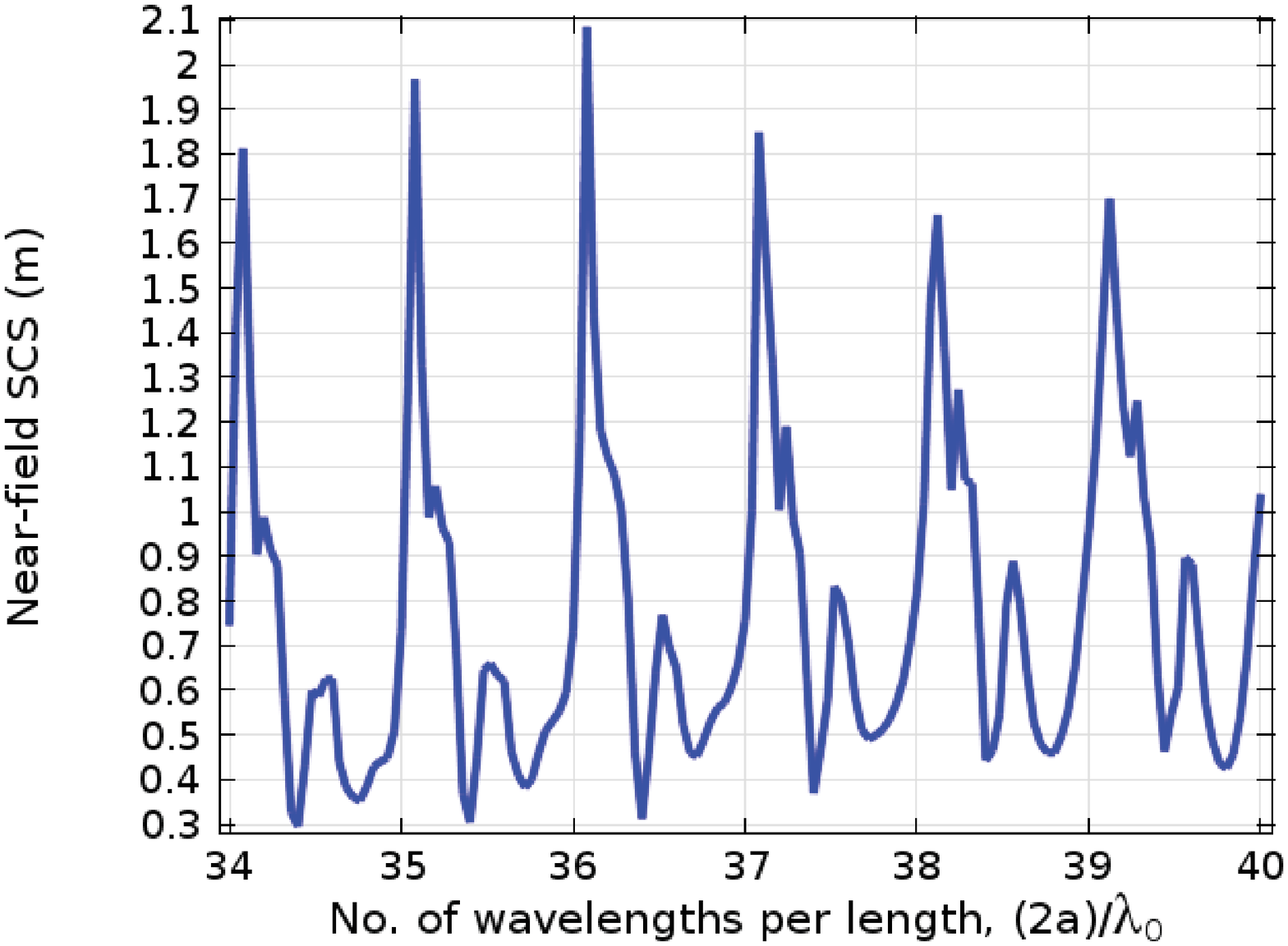}\\
(a)&(b)\\
\end{tabular}
\caption{(color online). Directional TM-polarization cloak without magnetic response ($\mu_{zz}=1$):
(a) Field plot ($H_z$) at $\nu=5.396$~GHz.
(b) Visibility figure of merit ($\sigma_{NF}$) as a function of frequency, showing a series of FPR-like transmission resonances.
}
\label{fig:cloak_impedance_scaled}
\end{figure}

The subluminal cloak with $\epsilon\ge1$ (Fig.~\ref{fig:cloak_index_scaled}) still requires $\mu>1$. While a medium with $\mu>1$ can theoretically have arbitrarily small dispersion and negligible loss, such properties have only been found in materials at low frequencies.
To tackle the magnetic response issue, we recall that in the geometric optics (GO) limit, the scaling of the wave impedance does not change ray trajectories. Arbitrary scaling of the impedance profile was used by various authors and is known as the eikonal limit~\cite{schurig_smith06,urzhumov_smith_njp10,urzhumov_smith_prl10,urzhumov_pendry11} of TO. However, previous works considered only continuous impedance scaling, with the idea that, even in the GO limit, impedance discontinuities create reflections. Omission of reflected rays on impedance-mismatched interfaces led some authors to believe that CM-based cloaks may perform well at arbitrary incidence angles~\cite{leonhardt06,urzhumov_landy_smith_jap12}.

While impedance discontinuities would increase visibility, we may still take advantage of the quasi-one dimensional geometry of our structure. In 1D, perfect invisibility can be easily achieved --- though only for a discrete set of $\lambda_0$ --- using the well-known phenomenon of FPR. A lossless homogeneous slab exhibits 100\% transmittance at certain $\lambda_0$, regardless how large or small the impedance mismatch. Although a rectangular cloak is not exactly a 1D object, we may expect that FPR-like resonances could strongly reduce the forward-scattering amplitude (i.e., the shadow). By virtue of the 2D optical theorem~\cite{boya_murray94}, then, the total extinction cross-section is also strongly suppressed.

Full-wave simulations of the original (index-unscaled) cloak whose impedance was adjusted to have $\mu_{zz}=1$ in all regions (Fig.~\ref{fig:cloak_impedance_scaled}) demonstrate that, indeed, numerous such resonances exist, and they strongly reduce visibility of the structure at select $\lambda_0$. These resonances can be utilized as long as the length of the cloak $2a$ exceeds $\lambda_0/2$. Their abundance in the GO limit, $\lambda_0/a\to 0$, makes it easy to choose a close-to-desirable $\lambda_0/a$ ratio of the cloak.
We emphasize that the use of either approximation presented here creates a fundamental bandwidth limitation for the cloak --- a trade-off to eliminating dispersive and lossy media.

%



This work was supported through a Multidisciplinary University Research Initiative, sponsored by the U.S. Army Research Office (Grant No. W911NF-09-1-0539), and partially supported by the U.S. Navy through a subcontract with SwampWorks (Contract No. N00167-11-P-0292). The authors are grateful to Nathan Landy (Duke University) for useful discussions on feasibility of directional cloaks.


%
%

\pagebreak
\section*{Informational Fourth Page}

%
%
%

\end{document}